\documentclass[
 preprint,
 amsmath,amssymb, aps,
 superscriptaddress
]{revtex4-2}
\usepackage{graphicx}
\usepackage{dcolumn}
\usepackage{bm}

\begin{document}

\preprint{APS/123-QED}

\author{Dongmin Seo}
\affiliation{
 Department of Physics, University of Seoul, Seoul 02504, Republic of Korea 
 }
\affiliation{
 Department of Smart Cities, University of Seoul, Seoul 02504, Republic of Korea
 }

 \author{Gihyeon Ahn}
\affiliation{
Department of Physics, Hanyang University, Seoul 04763, Republic of Korea
}
\author{Gaurab Rimal}
 \affiliation{
 Department of Physics and Astronomy, Rutgers, The State University of New Jersey, Piscataway, New Jersey 08854, USA
}
\author{Seunghyun Khim}
\affiliation{%
Max Planck Institute for Chemical Physics of Solids, Nöthnitzer Straße 40, 01187 Dresden, Germany
}
\author{Suk Bum Chung}
\affiliation{
 Department of Physics, University of Seoul, Seoul 02504, Republic of Korea 
 }
\affiliation{
 Natural Science Research Institute, University of Seoul, Seoul, 02504, Republic of Korea
 }
\affiliation{
 School of Physics, Korea Institute for Advanced Study, Seoul, 02455, Republic of Korea
 }
\author{A. P. Mackenzie}
\affiliation{%
Max Planck Institute for Chemical Physics of Solids, Nöthnitzer Straße 40, 01187 Dresden, Germany
}%
\affiliation{
 Scottish Universities Physics Alliance, School of Physics and Astronomy,
University of St Andrews, St Andrews KY16 9SS, United Kingdom
}
\author{Seongshik Oh}
\affiliation{
 Department of Physics and Astronomy, Rutgers, The State University of New Jersey, Piscataway, New Jersey 08854, USA
}%
\author{S. J. Moon}
 \email{soonjmoon@hanyang.ac.kr}
\affiliation{
Department of Physics, Hanyang University, Seoul 04763, Republic of Korea
}
\thanks{Corresponding author}
\author{Eunjip Choi}
 \email{echoi@uos.ac.kr}
\affiliation{
 Department of Physics, University of Seoul, Seoul 02504, Republic of Korea 
 }
\thanks{Corresponding author}


\date{\today}

\title{Interaction of in-plane Drude carrier with c-axis phonon in $\rm PdCoO_2$}

\date{\today}

\begin{abstract}

We performed polarized reflection and transmission measurements on the layered conducting oxide $\rm PdCoO_2$ thin films.
For the ab-plane, an optical peak near   $\Omega$ $\approx$ 750 cm$^{-1}$ drives  
the scattering rate $\gamma^{*}(\omega)$ and effective mass $m^{*}(\omega)$  of the Drude carrier to  increase and decrease respectively for $\omega$ $\geqq$ $\Omega$. 
For the c-axis,
a longitudinal optical phonon (LO) is present at $\Omega$ as evidenced by a peak in the loss function Im[$-1/\varepsilon_{c}(\omega)$]. 
Further polarized measurements in different light propagation (q) and electric field (E) configurations indicate that 
the Peak at $\Omega$ results from an electron-phonon coupling of the ab-plane carrier with the c-LO phonon, which leads to the frequency-dependent $\gamma^{*}(\omega)$ and  $m^{*}(\omega)$. 
This unusual interaction was previously reported in high-temperature superconductors (HTSC) between a non-Drude, mid-infrared band and a c-LO. On the contrary, 
it is the Drude carrier that couples in $\rm PdCoO_2$.  The coupling between the ab-plane Drude carrier and c-LO suggests that the c-LO phonon may play a significant role in the characteristic ab-plane electronic properties of $\rm PdCoO_2$ including the ultra-high dc-conductivity, phonon-drag, and hydrodynamic electron transport.

\end{abstract}

\maketitle

{\it Introduction}.---The interaction of an electron with a phonon plays a key role in emergent phenomena such as the polaron, charge density wave, and superconductivity. 
The electron-phonon interaction manifests itself, among others, in the ac-response of the material, including optical reflectance and dielectric functions \cite{basov2011electrodynamics}. 
In HTSC, the ab-plane optical conductivity exhibits an electronic continuum at the mid-IR range.  Interestingly, for most HTSC compounds such as $\rm YBa_{2}Cu_{3}O_{7-\delta}$ \cite{kamaras1990clean}, $\rm Bi_2Sr_2CaCu_2O_8$ \cite{reedyk1988far}, and others \cite{reedyk1992optical, foster1990infrared}, 
a particular type of spectral feature, i.e., narrow dips or minima appear on top of the broad mid-IR band at multiple photon energies. 
In 1992, Reedyk and Timusk discovered that the minima are associated with optical phonons propagating along the c-axis of the lattice, specifically, longitudinal optical phonons. The unusual activation of the c-axis phonons in the ab-plane reflectivity, 
normally forbidden due to the momentum selection rule,  
results from the coupling of the in-plane electron with the c-axis LO phonons \cite{reedyk1992evidence}. This electron-phonon interaction has drawn attention from the perspective of possible superconductivity pairing mechanisms. On the other hand, there has been a question as to whether a similar kind of interaction occurs in other layered metallic oxides as well. To the best of our knowledge, such material has not yet been reported to date. 

 The delafossite $\rm PdCoO_2$ consists of triangular Pd-planes that alternate with the $\rm CoO_6$ planes and are stacked along the c-axis. The in-plane electrical conduction occurs predominantly in the Pd-sheet \cite{higuchi1998photoemission, seshadri1998metal, hasegawa2001electronic, eyert2008metallic, noh2009anisotropic, noh2009orbital, ong2010origin,mackenzie2017properties,daou2017unconventional,harada2021thin}, giving rise to dc-conductivity $\sigma$ = 3.8 $\times$ $10^{5}$ $\Omega^{-1}$cm$^{-1}$ at room temperature \cite{hicks2012quantum} which is, remarkably, higher than noble metals such as Au or Ag \cite{lynch1997comments}. The mean free path of electrons is as long as 20 $\mu$m at low-T, making this material a promising candidate for the hydrodynamic and other non-local transport studies \cite{moll2016evidence}. 
An optical study by Homes {\it et al.} suggested that, importantly, the  
ab-plane electrons may couple with c-axis LO phonons in $\rm PdCoO_2$ \cite{homes2019perfect}. This claim was based on two phonon-like peaks that are expected to be silent in the ab-plane reflectivity, yet appear in the actual measurements. This interesting suggestion, however, was not supported by compelling experimental evidence.  

In this work, we directly address this issue by performing optical measurements using a distinct approach from Ref.\cite{homes2019perfect}: Firstly, we probe both the ab-plane and the c-axis. For the ab-plane study, we employ a thin film $\rm PdCoO_2$ instead of a single crystal. The latter has an extremely high reflection in the infrared range, which, as mentioned in Ref.\cite{homes2019perfect}, poses difficulty in carrying out a quantitative analysis. Such a problem can be largely alleviated by using a $\rm PdCoO_2$ thin film for which reflectivity is significantly reduced. Additionally, a thin film allows for transmission measurements, which, when combined with the reflection, leads to precise optical dielectric functions. Secondly, for the c-axis study, we take advantage of a single crystal $\rm PdCoO_2$ in combination with a focused beam of microscopic FTIR,  which makes the optical measurement possible despite the limited sample dimension along the c-axis.
Through the complementary studies on a thin film (for the ab-plane) and a single crystal (for the c-axis), we firmly establish that the ab-plane electrons of $\rm PdCoO_2$ couple with a longitudinal c-axis optical phonon.
While the coupling in HTSC occurred between the (non-Drude) mid-infrared band and c-LO,  it is the Drude carrier that couples in $\rm PdCoO_2$.


{\it Experiment}.---
Epitaxial $\rm PdCoO_2$ thin films (thickness = 90 nm) were grown on an $\rm Al_2O_3$ substrate using the molecular beam epitaxy (MBE) technique and were thoroughly characterized through various methods
such as XRD, RHEED, TEM, etc. \cite{brahlek2019growth}. 
The ab-plane optical transmittance and reflectance in the infrared range were measured on the thin films using FTIR (Bruker Vertex 70v) in combination with the {\it in-situ} gold evaporation technique \cite{homes1993technique}.  A Spectroscopic Ellipsometer (J.A. Woollam VASE) was used to obtain the optical dielectric functions from 0.7 eV to 4 eV. The optical reflection of the c-axis was measured on a 100 $\mu$m-thick high-quality single crystal grown using the flux method \cite{shannon1971chemistry, takatsu2010unconventional}, in combination with microscopic FTIR (Hyperion 2000). 
The a and b directions refer to a set of two orthogonal directions in the hexagonal plane, which is not aligned with respect to the crystal structure.

\begin{figure}
\centering
\includegraphics[width=0.8\linewidth]{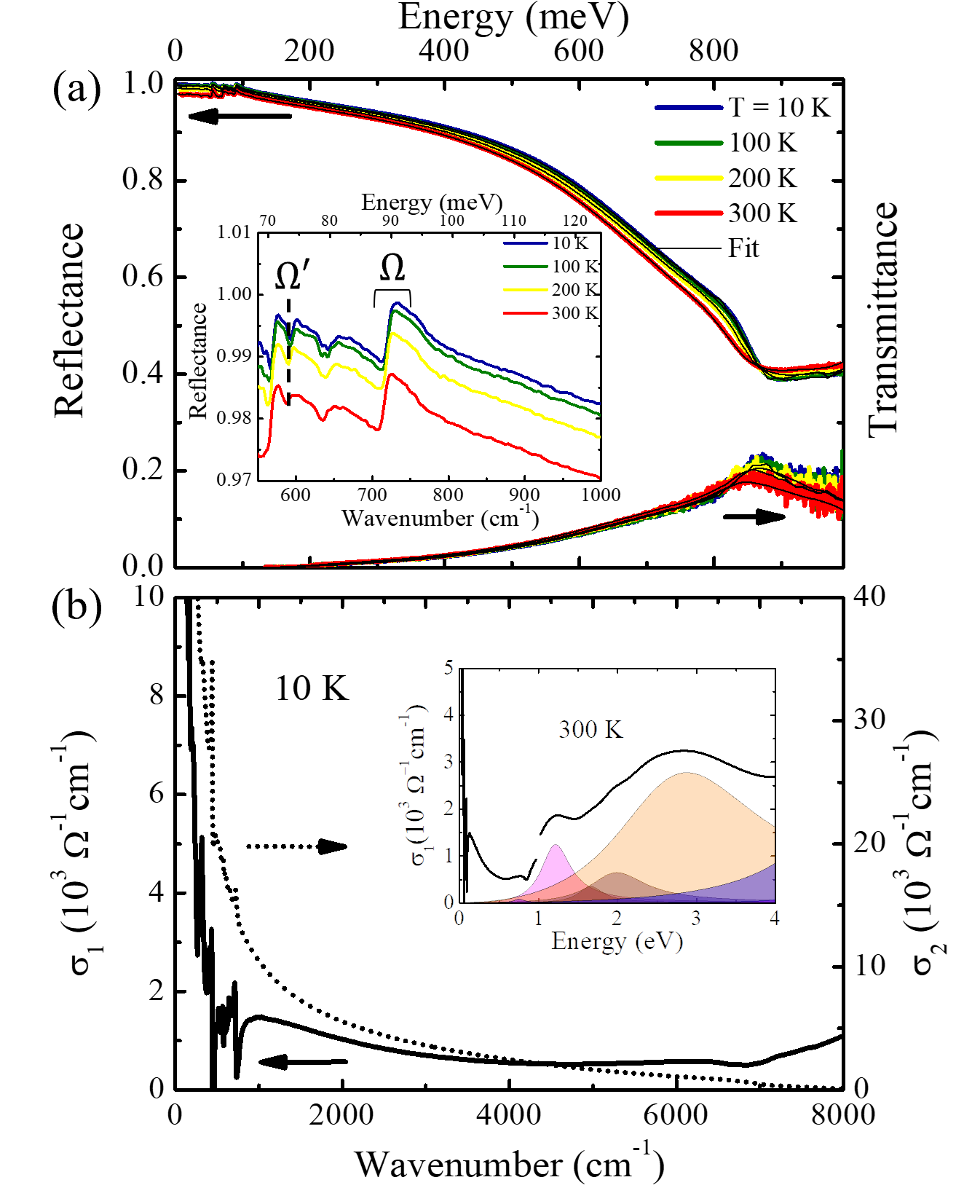}
\caption{(a) Reflectance and transmittance of a $\rm PdCoO_2$ thin film (thickness = 90 nm). The inset highlights that there is
a peak around $\sim$ 90 meV. (b) Real and imaginary parts of the optical conductivity. Inset depicts the wide-range $\sigma_1(\omega)$ up to 4 eV.} 
\label{fig1}
\end{figure}

{\it Results}.---Fig.~\ref{fig1} (a)  shows the reflectance $\rm R(\omega)$ and transmittance $\rm T(\omega)$
 of the $\rm PdCoO_2$ thin film for $\hbar\omega$ $<$ 0.1 eV. We fit $\rm R(\omega)$ and $\rm T(\omega)$ simultaneously using the multilayer (film+substrate) analysis algorithm of the Kramers-Kronig (KK) constrained RefFit program \cite{kuzmenko2005kramers, kuzmenko2016reffit}. The dielectric functions of the bare $\rm Al_2O_3$-substrate were characterized separately and fed into the analysis. 
  Fig.~\ref{fig1} (b) displays $\sigma_1(\omega)$ and $\sigma_2(\omega)$, the real and imaginary optical conductivity of $\rm PdCoO_2$, respectively, obtained from the fit at $T$ = 10 K. 
  They consist of an intra-band (Drude) response in the low-energy range and inter-band transitions at high energy $\hbar\omega$ $>$ 0.8 eV. When compared to the previous optical study on a single crystal $\rm PdCoO_2$ \cite{homes2019perfect}, the inter-band transitions of our film are almost identical, whereas the Drude peak is notably broader. The latter is attributed to additional scatterings of the carrier at the twin-boundary and the top / bottom surfaces of the film \cite{brahlek2019growth}. In the inset of Fig.~\ref{fig1} (a) we highlight that there is a distinct peak-like feature at $\hbar\omega$ = 90 meV 
  in $\rm R(\omega)$. We label it conveniently as Peak-$\Omega$ and will revisit it frequently later for data analysis. To add, $\Omega^{\prime}$ refers to a dip at a lower energy. 
  
\begin{figure}
\centering
\includegraphics[width=0.8\linewidth]{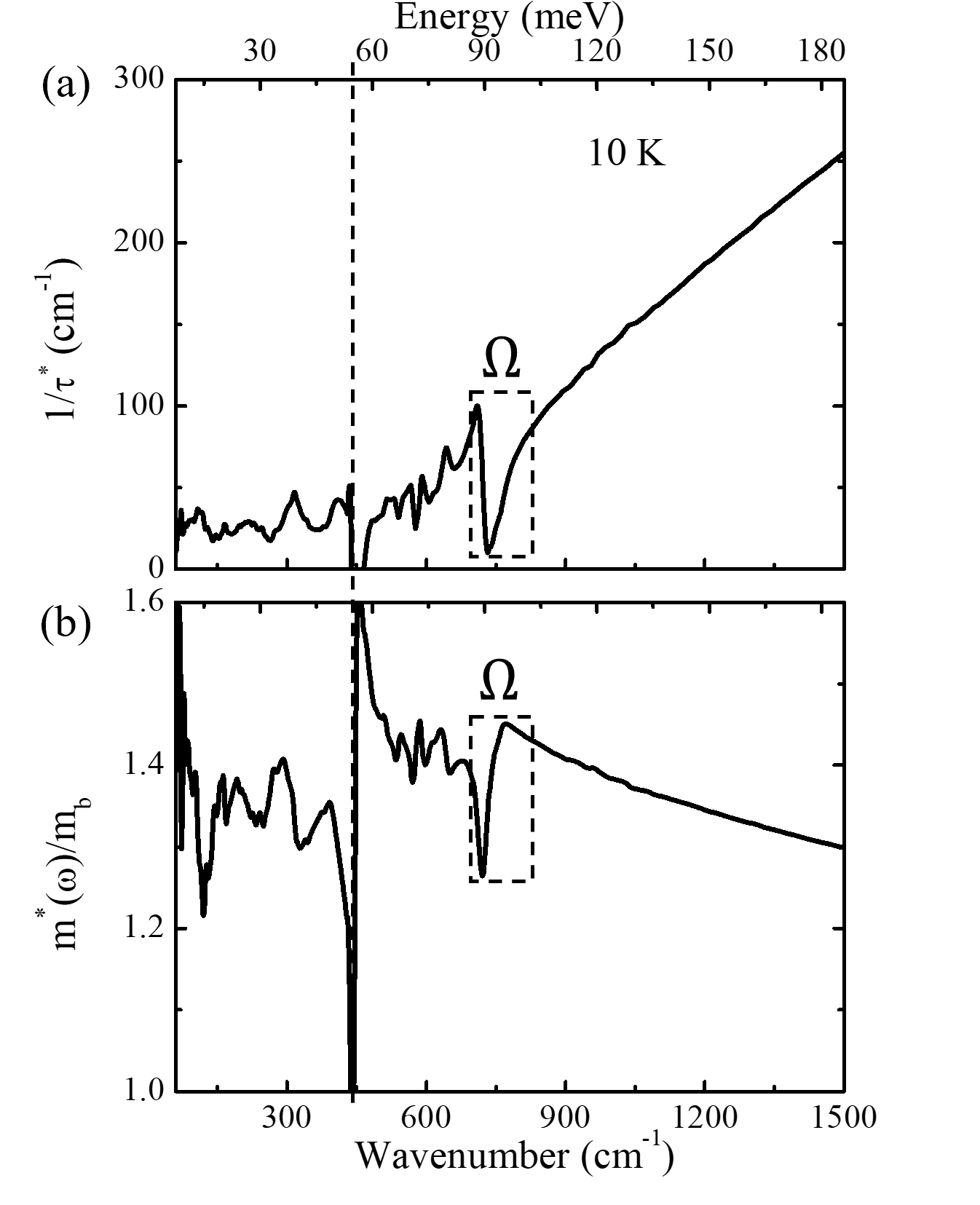}
\caption{ Frequency-dependent scattering rate $1/\tau^{*}(\omega)$ and mass enhancement $m^{*}(\omega)/m_b$. They are calculated from the Drude conductivity through Eq. (1) and Eq. (2).  Here, $\Omega$ corresponds to the optical feature in the reflectivity, Peak-$\Omega$ in Fig. 1(a). The spurious noise at $\omega$ = 450 cm$^{-1}$ (dashed line) is caused by the substrate. (See Fig. S2) }
\label{fig2}
\end{figure}
In Fig.~\ref{fig2} we show the scattering rate $1/\tau^{*}(\omega)$ 
and effective mass $m^{*}(\omega)$ of the Drude carrier. They are calculated from the Drude
$\sigma_{1}(\omega)$ and $\sigma_{2}(\omega)$ using the extended Drude analysis formula 

\begin{equation} 
\frac{1}{\tau^{*}(\omega)}=
\frac{\omega \sigma_1(\omega)}{\sigma_2(\omega)}
\label{eq1-a}
\end{equation}

\begin{equation} 
\frac{m^*(\omega)}{m_b}=
\frac{\omega_p^2}{4\pi}\frac{\sigma_2(\omega)}{\sigma_1^2(\omega)+\sigma_2^2(\omega)}\frac{1}{\omega}
\label{eq1-b}
\end{equation}

The $1/\tau^{*}(\omega)$ increases markedly for $\omega$ $>$ $\Omega$ and $m^{*}(\omega)$ drops from the same frequency.
At $\omega$ = 90 meV, there is a dispersive structure in $1/\tau^{*}(\omega)$ and $m^{*}(\omega)$ that triggers the frequency-dependent changes. This structure originates from the Peak-$\Omega$
in $\rm R(\omega)$. 
The frequency-dependent  $1/\tau^{*}(\omega)$ and $m^{*}(\omega)$ are the characteristic behavior of an electron-boson interaction: For a conducting material with an electron-boson interaction,  $1/\tau^{*}(\omega)$ increases as $\omega$ exceeds the boson energy, and simultaneously, $m^{*}(\omega)/m_b$ begins to decrease from a dressed mass ($>$ 1) to bare band mass (= 1) \cite{carbotte1990properties,allen1971electron,shulga1991electronic,  stricker2014optical,kostic1998non,schlesinger1990superconducting}. Fig.~\ref{fig2} shows that an electron-boson coupling is occurring in $\rm PdCoO_2$ with a boson mode located at $\omega$ = $\Omega$.

To ensure that the frequency-dependent $1/\tau^{*}(\omega)$ and $m^{*}(\omega)$ are intrinsic properties of $\rm PdCoO_2$, we synthesized a pure Pd-thin film (thickness = 15 nm) using MBE on the same substrate ($\rm Al_2O_3$) and performed the same optical measurements (R and T) and extended-Drude analysis. The Pd-film is classified as a noble metal such as Au or Ag films. In contrast to $\rm PdCoO_2$,
$1/\tau^{*}(\omega)$ and $m^{*}(\omega)$ of the Pd-film are independent of frequency as expected for a simple Drude metal. (See Supplementary Fig. S1) This comparative study supports that $1/\tau^{*}(\omega)$ and $m^{*}(\omega)$ in Fig. 2
are not artifacts caused by, for example, the $\rm Al_2O_3$ substrate but
are the genuine properties of $\rm PdCoO_2$. In addition, the Pd-film does not show Peak-$\Omega$ (and $\Omega^{\prime}$) in R($\omega$) indicating that Peak-$\Omega$ (and $\Omega^{\prime}$) is intrinsic to $\rm PdCoO_2$ as well. (See Supplementary Fig. S2).
In Fig. 2, 
$1/\tau^{*}(\omega)$ and $m^{*}(\omega)$ become considerably noisy for the range $\omega <$ $\Omega$. The unwanted noises are caused mostly by the strong optical phonons of $\rm Al_2O_3$, which we present in detail in Supplementary Fig. S2. 
Ideally, the substrate phonons should be isolated from the thin film during the data analysis (fitting). However, in practice, they are not perfectly removed, causing the noise.
Importantly, Fig. S2 demonstrates that, unambiguously,  Peak-$\Omega$  arises not from
the substrate but from $\rm PdCoO_2$. 
 Given that Peak-$\Omega$ is driving the frequency-dependent $1/\tau^{*}(\omega)$ and $m^{*}(\omega)$, it is crucial to unveil the origin of Peak-$\Omega$ in order to understand the electron-boson interaction of $\rm PdCoO_2$. 

\begin{figure}[h!]
\centering
\includegraphics[width=0.8\linewidth]{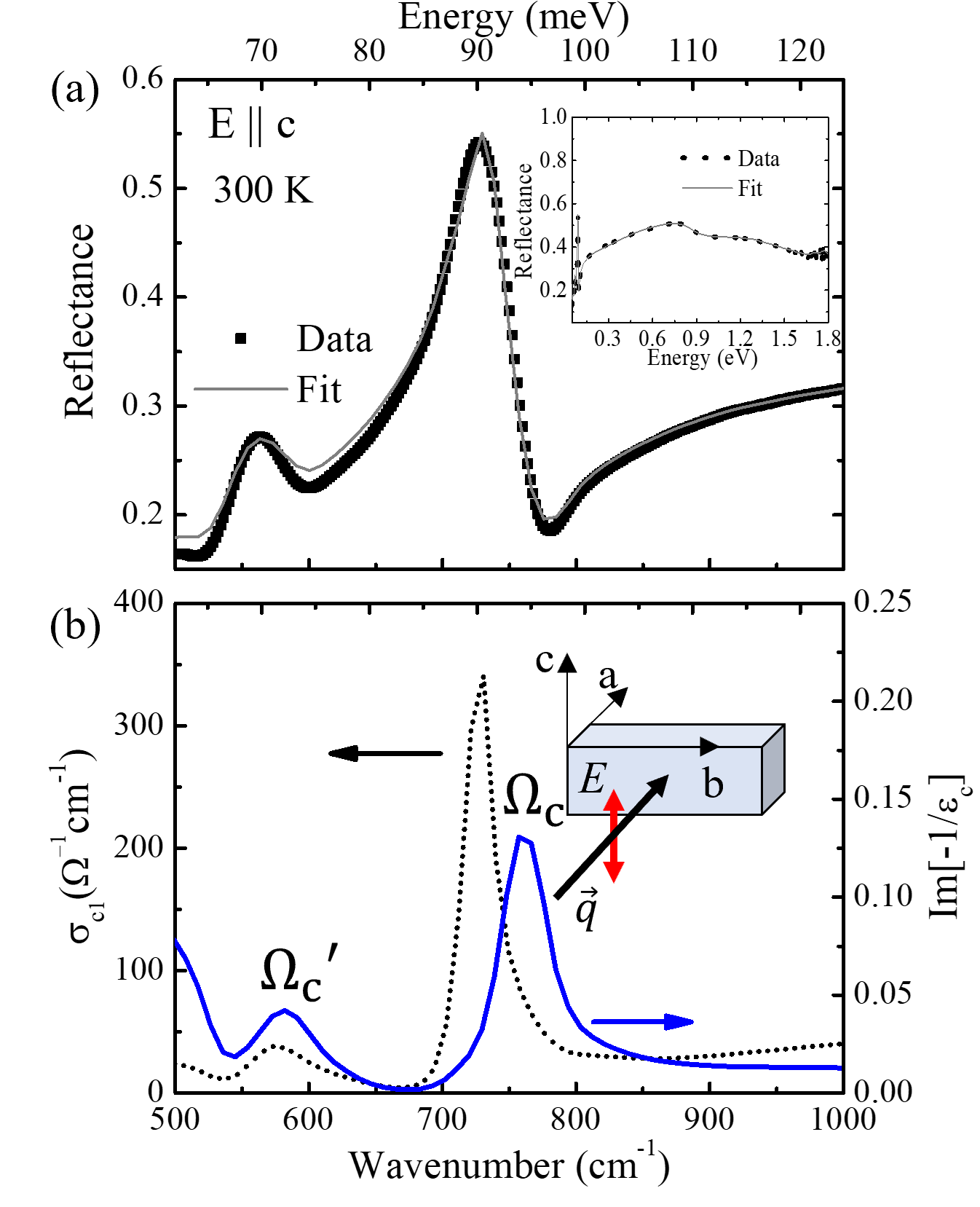}
\caption{
(a) Reflectance $\rm R_{c}(\omega)$ measured with the light polarized as $E$ $\|$ c on the $\rm PdCoO_2$ single crystal. Inset shows the wide-range $\rm R_{c}(\omega)$ up to 1.8 eV. (b) The c-axis optical conductivity and dielectric loss function. Inset shows that the light propagates along the ab-plane, $q$ $\|$ ab, and E-field is polarized along the c-axis, $E$ $\|$ c. Here $\Omega_{C}$ and $\Omega_{C}^{\prime}$ denote the two peaks of the Im[$-1/\varepsilon_c(\omega)$].
}
\label{fig3}
\end{figure}
In Fig.~\ref{fig3} (a) we measured the c-axis reflectance $\rm R_{c}(\omega)$ of the $\rm PdCoO_{2}$ single crystal. To measure $\rm R_{c}(\omega)$,
a focused IR beam from a microscopic FTIR polarized along the c-axis ($E$ $\|$ c) was illuminated on the side facet of the d $\sim$ 100 $\mu$m-thick single crystal. In this manner, reproducible data were obtained for $\omega$ $>$ $\sim$ 500 cm$^{-1}$. 
The $\rm R_{c}(\omega)$ shows a prominent structure at $\hbar\omega$ = 90 meV and a minor one at 70 meV.
The wide-range $\rm R_{c}(\omega)$ (inset) reveals an insulating behavior of the c-axis, which contrasts sharply with the metallic $\rm R(\omega)$ of the ab-plane.
We fit the $\rm R_{c}(\omega)$ using the KK-constrained RefFit and calculated the complex c-axis optical conductivity $\sigma_{c}(\omega)$ and dielectric constant $\varepsilon_{c}(\omega)$. 
 In the fit, we constrained $\sigma_{c}(\omega)$ to match the c-axis dc-conductivity at $\omega$ = 0 \cite{putzke2020h}.
In Fig.~\ref{fig3} (b), we show the real part of $\sigma_{c}(\omega)$. $\sigma_{c1}(\omega)$ is peaked at $\hbar\omega$ = 730 cm$^{-1}$ and 570 cm$^{-1}$, which represents two transverse optical phonons (TO) of the c-axis. These c-TO phonons propagate along the ab-plane, q $\|$ ab.
We also calculate the dielectric loss function Im[$-1/\varepsilon_{c}(\omega)$], which shows two peaks $\Omega_{c}$ and $\Omega_{c}^{\prime}$ representing the c-axis LO phonons. They propagate along q $\|$ c.  Note that, remarkably, $\Omega_{c}$  is very close to Peak-$\Omega$, suggesting that it is a possible source of Peak-$\Omega$. The c-TO phonon (q $\|$ ab), which is also close to Peak-$\Omega$,  is not excited in the normal-incidence thin film measurements (q $\|$ c) in Fig. 1, thus cannot create Peak-$\Omega$. As for $\Omega_{c}^{\prime}$, its energy is close to the dip $\Omega^{\prime}$ of Fig. 1.  
\newline

\begin{figure}[h!]
\centering
\includegraphics[width=0.8\linewidth]{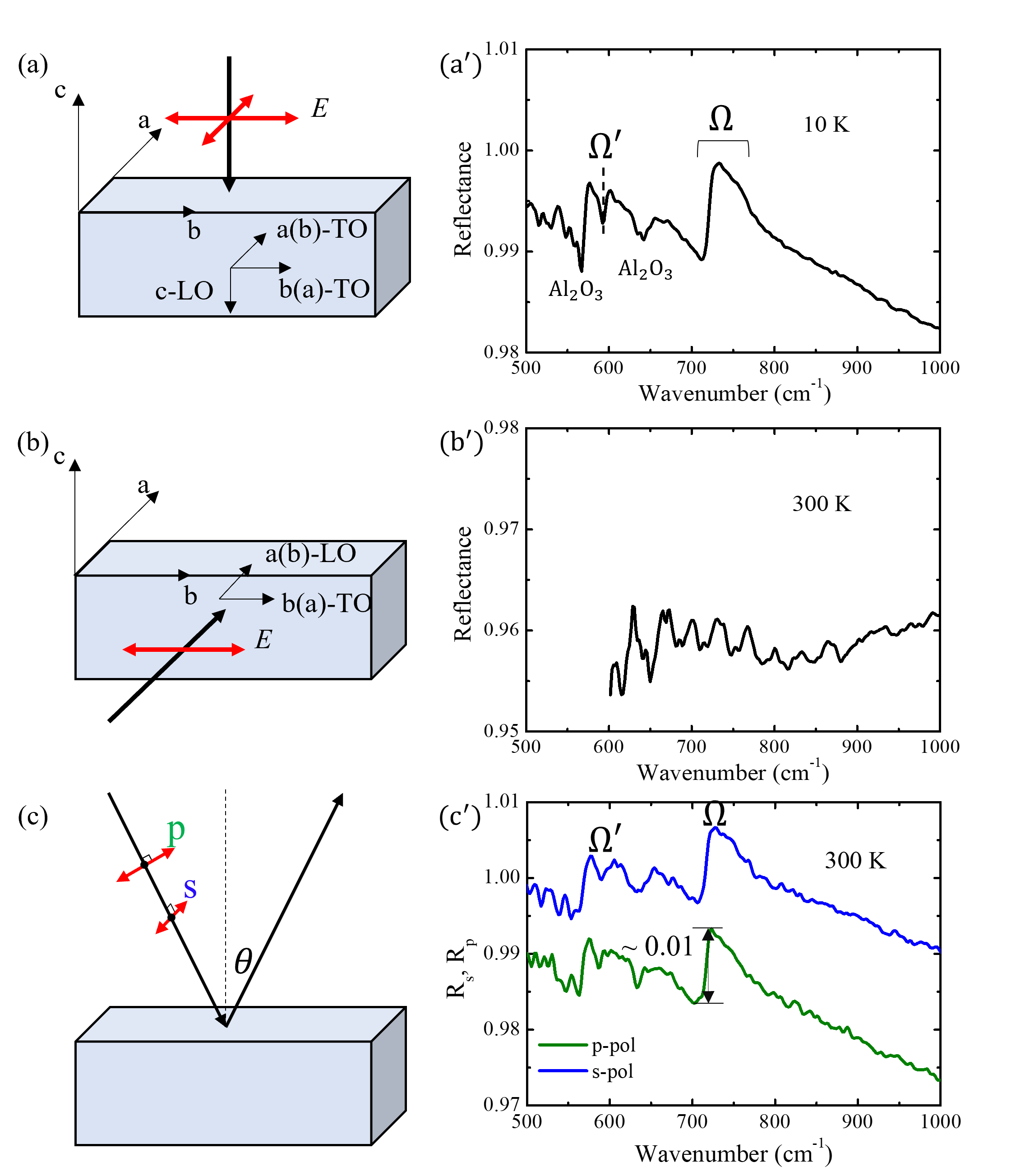}
\caption{
Polarization-dependent reflectance of $\rm PdCoO_2$. (a) light propagates along q $\|$ c and $E$-field is unpolarized. (b) light propagates along $q$ $\|$ a(b) and $E$-field is polarized along $E$ $\|$ b(a). 
(c) s- and p-polarized lights are incident at an incidence angle $\theta$ = 10$^{\circ}$. The reflectance data of (a), (b), and (c) are shown in (a$^{\prime}$), (b$^{\prime}$), and (c$^{\prime}$), respectively. The thin film was used for 
(a) and (c), and the single crystal was used for (b).
}
\label{fig4}
\end{figure}


To confirm the presumption that Peak-$\Omega$ originates from  $\Omega_{c}$, we perform further polarized reflectance measurements.
In Fig.~\ref{fig4} (a) incident light propagates along the c-axis while the electric field is parallel to the ab-plane. This optical configuration ($q$ $\|$ c, $E$ $\|$ ab) can activate the ab-plane TO and the c-axis LO phonon. 
In Fig.~\ref{fig4} (b) we employed a different optical configuration $q$ $\|$ a and $E$ $\|$ b which activates the b-TO but not the c-LO. (To note, we use 'a' and 'b' to represent two orthogonal axes of the ab-plane but they do not indicate  any specific crystallographic directions. The terms a-TO and b-TO are equivalent to ab-TO.) The R($\omega$) in Fig.~\ref{fig4} (a$^{\prime}$) and (b$^{\prime}$) show that Peak-$\Omega$ is activated in Fig.~\ref{fig4} (a), but is absent in Fig.~\ref{fig4} (b), demonstrating that ab-TO is excluded from the source of Peak-$\Omega$, thus leaving the c-LO ($\Omega_{c}$) the only remaining candidate. The ab-TO phonons are another possible sources of  Peak-$\Omega$ but, according to Ref. \cite{homes2019perfect}, they are far from Peak-$\Omega$. 
In general, a c-axis optical phonon of a layered material does not appear in the ab-plane reflectivity due to forbidden symmetry. 
In $\rm PdCoO_2$, however, $\Omega_{c}$ manifests itself in the ab-plane reflectivity as a result of coupling with the ab-plane Drude carrier. This coupling leads to the frequency-dependent $1/\tau^{*}(\omega)$ and $m^{*}(\omega)$ of Fig. 2. 

In our near-normal ($\theta$ = 10$^{\circ}$) reflectance measurement, incident light contains a small $E$ $\|$ c 
component, which may cause the c-axis phonons to leak into the ab-plane reflectivity. In this case, Peak-$\Omega$ may appear in R($\omega$) even if the electron-phonon coupling were absent.  
To test if this is the case for Fig.~\ref{fig4} (a$^{\prime}$), we measured R($\omega$) using the s- and p-polarization as shown in Fig.~\ref{fig4} (c): In the s-polarization, the light has no $E$ $\|$ c component, whereas the p-polarization does have a finite $E$ $\|$ c component. The R($\omega$) in Fig.~\ref{fig4} (c$^{\prime}$) shows that Peak-$\Omega$ is activated in the s-polarization with similar strength as in the p-polarization. This result rules out the leakage scenario of Peak-$\Omega$.
To reinforce our conclusion, we theoretically the calculated grazing-incidence R($\omega$) 
at incidence angles 
$\theta$ = 10$^{\circ}$ and 20$^{\circ}$. For this, we used  
the ab-plane and c-axis dielectric functions measured in Fig.~\ref{fig1} and Fig.~\ref{fig3}, respectively. 
The calculation results, shown in the  Supplementary   
Fig. S3, reveal that at $\theta$ = 20$^{\circ}$,  the c-LO 
leaks into the ab-plane reflectivity in the p-polarization, giving rise to a peak with 5$\times 10^{-4}$ in height.  However, this peak height is far weaker than the actual height of Peak-$\Omega$ in Fig.~\ref{fig4} (c$^{\prime}$), 0.01. Furthermore, at the experimental angle $\theta$ = 10$^{\circ}$,
the calculated leakage becomes even smaller, and the peak is too weak to be detected. 
This observation  supports  again that the $E$-field leakage cannot account for Peak-$\Omega$  in $\rm R(\omega)$. We thus conclude that the c-LO does couple with the ab-plane Drude carrier manifesting itself as Peak-$\Omega$ in the ab-plane reflectance. 
\newline

{\it Discussion}.---To compare $\rm PdCoO_2$ with HTSC, they are the two types of rare materials that exhibit the coupling of the ab-plane carrier with c-LO phonons. One major difference, however, is that 
the Drude carrier couples in $\rm PdCoO_2$  whereas it is the mid-IR band in HTSC \cite{reedyk1992evidence}. 
Therefore, in the latter, c-LO does  not influence the dc-transport. On the contrary,  the c-LO of $\rm PdCoO_2$ may play a significant role in the ab-plane transport such as the hydrodynamic charge flow.  
We emphasize that $\rm PdCoO_2$ is the first layered material in which the c-LO couples with the ab-plane Drude carriers.


To discuss $\Omega^{\prime}$, we examine if it arises from $\Omega_{C}^{\prime}$ like $\Omega$ did from $\Omega_{C}$. For this, we compare the ab-plane $\sigma_{1}(\omega)$ with the c-axis Im[$-1/\varepsilon_{c}(\omega)$] in Fig. S4 following a similar approach as in Ref.\cite{reedyk1992evidence}.  The Peak-$\Omega^{\prime}$ occurs at the same energy as $\Omega_{C}^{\prime}$ but with a significantly narrower width.  
On the contrary, for the single crystal $\rm PdCoO_2$,  an optical peak occurs in the ab-plane
at the same energy and with similar width as  $\Omega_{C}^{\prime}$, supporting that the electron-phonon interaction persists for $\Omega^{\prime}$ \cite{homes2019perfect}.
In the thin film, $\sigma_{1}(\omega)$ is highly uncertain in the region of $\Omega^{\prime}$, hindering precise determination of the spectral shape. 
To definitely establish the correlation with $\Omega_{C}^{\prime}$,
improved measurements that overcome the noise are needed. 
To further compare the thin film and single crystal results, we note that Peak-$\Omega$ has  
an asymmetric, Fano-like shape in $\sigma_{1}(\omega)$ in both cases. However, they have the opposite Fano-asymmetry signs and different  strengths of asymmetry (See Supplementary Fig. S4  and Ref. \cite{homes2019perfect}). Such  differences suggest that substantial thickness-dependent effects exist for the electron-phonon coupling.
Lastly, while we focused primarily on electron-phonon coupling in this paper,  our data may suggest that another type of interaction, such as the electron-electron interaction, may apply to $\rm PdCoO_2$ as well. (See Supplementary Fig. S5)

{\it Conclusion}.---In conclusion, we performed polarized infrared transmission and reflection measurements on a $\rm PdCoO_2$ thin film. 
In the ab-plane, the scattering rate $\gamma^*(\omega)$ and effective mass $m^*(\omega)$ of the Drude carriers increased and decreased for $\omega > \Omega$, respectively,  driven by Peak-$\Omega$.
In the c-axis measurement on a single crystal, 
a longitudinal optical phonon was found at $\Omega_{C}$ as evidenced by a peak of the loss function Im[$-1/\varepsilon_{c}(\omega)$]. 
Further optical measurements  employing  different (q, E) configurations revealed  that 
Peak-$\Omega$ is activated due to the interaction of the ab-plane Drude carriers with the c-LO phonon. This electron-phonon interaction leads to the frequency-dependent $\gamma^{*}(\omega)$ and  $m^{*}(\omega)$.
Our conclusion was established through the extensive supporting measurements on the pure Pd-film, bare $\rm Al_2O_3$ substrate, and the s- and p-polarized grazing-incidence reflection calculations. 
The coupling of the ab-plane Drude electron with the c-LO phonon implies that c-LO may play a significant role in the characteristic ab-plane carrier dynamics of $\rm PdCoO_2$ such as the ultra-high dc-conductivity, phonon-drag, and hydrodynamic charge flow, which is worthy for further studies.
We want to remark that we were  heavily assisted by Ref.\cite{reedyk1992evidence} and Ref.\cite{homes2019perfect} in conducting data analysis and interpretations.



{\it Acknowledgement}.---We thank C. C. Homes, D. Valentinis, J. Zaannen, D. Van Der Marel, Hyoung Joon Choi, and Han-Jin Noh for their helpful discussions. This work was supported by the NRF-2021R1A2C1009073 of Korea funded by the Ministry of Education.
D. S was partially supported by MOLIT as Innovative Talent Education Program for Smart City.
The work at Rutgers University is supported by National Science Foundation’s DMR2004125 and Army Research Office’s W911NF2010108.
S.B.C. was supported by the NRF of Korea funded by MSIT (2020R1A2C1007554) and the Ministry of Education (2018R1A6A1A06024977).
Research in Dresden benefits from the environment provided by the DFG Cluster of Excellence ct.qmat (EXC 2147, project ID 390858940).
The work at HYU was supported by the National Research Foundation grant of Korea (NRF) funded by the Korean government (MSIT) (2022R1F1A1072865 and RS-2022-00143178) and BrainLink program funded by the Ministry of Science and ICT through the National Research Foundation of Korea (2022H1D3A3A01077468). 

\bibliographystyle{apsrev4-2}
\bibliography{main}

\end{document}